\documentclass[aps,onecolumn,showpacs,nofootinbib,showkeys]{revtex4-2}
\usepackage[margin=2.5 cm]{geometry}

\RequirePackage[T1]{fontenc}
 
\usepackage{epsfig,graphicx,amsmath,amssymb,bm}
\RequirePackage{mathptmx}      
\usepackage{mathrsfs}
\usepackage{amssymb}
\RequirePackage{color}
\usepackage{changes}
\usepackage{slashed}
\usepackage{multirow}
\usepackage{scalerel}
\usepackage{tikz-feynman}
\usepackage{textcomp}
\usepackage{subcaption}
\usepackage[english]{babel}
\usepackage{float}
\RequirePackage{hyperref}
\hypersetup{
    linktocpage,
    colorlinks,
    citecolor=purple,
    filecolor=black,
    linkcolor=brown,
    urlcolor=purple,
}
\usepackage{nccmath}

\begin{document}

\title{
The decay $\tau \to 3\pi\nu_\tau$ and axial-vector meson $a_1$ in the NJL model
}

\author{Mikhail K. Volkov$^{1}$}\email{volkov@theor.jinr.ru}
\author{Kanat Nurlan$^{1,2,3}$}\email{nurlan@theor.jinr.ru}
\author{Aleksey A. Pivovarov$^{1}$}\email{tex_k@mail.ru}

\affiliation{$^1$ Bogoliubov Laboratory of Theoretical Physics, JINR, 
                 141980 Dubna, Russia \\
                $^2$ The Institute of Nuclear Physics, Almaty, 050032, Kazakhstan\\
                $^3$ L. N. Gumilyov Eurasian National University, Astana, 010008, Kazakhstan}   


\begin{abstract}
The branching fractions of $\tau \to \pi^+ \pi^-\pi^- \nu_\tau$ and $\tau \to \pi^- 2\pi^0\nu_\tau$ are calculated within the chiral NJL model. Features of the axial-vector $a_1$ meson which plays an important role in describing the $\tau$ decays are discussed. Permissible values for the mass and width of the $a_1$ meson are considered in accordance with the latest experiments.


\end{abstract}


\maketitle


The most probable decay of $\tau$ into three pseudoscalar mesons is the decay $\tau \to 3\pi \nu_\tau$ \cite{ParticleDataGroup:2022pth}.
Of particular interest to this process is the possibility of using it to study the internal properties of the axial-vector meson $a_1$ whose mass and width have not been well studied.
Indeed, the latest COMPASS collaboration experiment gave values for the mass and width $M_{a_{1}} = 1299 (+12, -28)$ MeV, $\Gamma_{a_{ 1}} = 380 \pm 80$ MeV \cite{COMPASS:2018uzl}. 
At the same time, the JPAC collaboration presented the values $M_{a_{1}} = 1209 \pm 4 (+12,-9)$ MeV, $\Gamma_{a_{1 }} = 576 \pm 11(+80, -20)$ MeV \cite{JPAC:2018zwp}. The determined value of the $a_1$ meson mass in these data differs by $3\sigma$. In the CLEO and LHCb experiments, axial-vector resonance structures in decays of $D^0$ meson were studied, where the parameters of the $a_1$ and $K_1$ states were measured \cite{dArgent:2017gzv, LHCb:2017swu}. It was found that the largest contribution to the processes $D^0 \to K^\mp \pi^\pm \pi^\pm \pi^\mp$ is came from axial-vector resonances \cite{LHCb:2017swu}. The above studies show that the determination of the axial-vector  $a_1$ meson mass and width is a topical issue.

A number of theoretical works have been devoted to this issue using various phenomenological models \cite{Girlanda:1999fu, GomezDumm:2003ku, Wagner:2008gz, Dumm:2009va, Nugent:2013hxa, Sadasivan:2021emk}. In these papers, the main attention was paid to obtaining data on the mass and width of the $a_1$ meson from the experimental decay width of $\tau \to 3\pi \nu_\tau$. In \cite{GomezDumm:2003ku, Dumm:2009va}, agreement with experiment was obtained at the meson mass $M_{a_1} = 1120$ MeV.

The Nambu-Jona-Lasinio (NJL) \cite{Nambu:1961tp, Ebert:1982pk, Ebert:1985kz, Volkov:1986zb, Vogl:1991qt, Klevansky:1992qe, Volkov:1993jw, Ebert:1994mf, Volkov:2005kw} model is a good tool for describing low-energy processes of meson interaction. In particular, this model turns out to be very useful in describing mesonic $\tau$ lepton decays \cite{Volkov:2017arr, Volkov:2022jfr}. 
One of the first attempts to use the NJL model to describe the $\tau \to 3\pi \nu_\tau$ process and obtain the parameters of the $a_1$ meson was made in the work \cite{Ivanov:1989qw} with the participation of one of the authors of this paper. An attempt was also made to determine the mass and width of the axial-vector meson $a_1$ using experimental data from the $\tau \to 3\pi \nu_\tau$ process. However, the experimental data on the decay of $\tau \to 3\pi \nu_\tau$ have changed markedly since then. Therefore, it is positively of interest to describe the decay of $\tau \to 3\pi \nu_\tau$ in the NJL model, taking into account the latest experimental data, which is the subject of the present paper.

The full interaction Lagrangians of mesons with quarks of the NJL model are given in \cite{Volkov:1986zb, Volkov:1993jw, Volkov:2005kw, Volkov:2017arr, Volkov:2022jfr}. The part of the Lagrangian for the strong interactions of $\pi$, $\rho$, and $a_1$ mesons with quarks necessary for describing the processes we are considering takes the form
\begin{eqnarray}
\label{L1}
\Delta{\mathcal L_{int}} & = & \bar{q}\left[i\gamma^{5} g_{\pi}\sum_{j = \pm,0}\lambda_{j}^{\pi}\pi^{j} 
+ \frac{1}{2} \gamma^{\mu} g_{\rho}\sum_{j = \pm,0}\lambda_{j}^{\rho}\rho_{\mu}^{j}
+ \frac{1}{2} \gamma^{\mu}\gamma^{5}g_{a_{1}}\sum_{j = \pm,0}\lambda_{j}^{\rho}a_{1\mu}^{j} \right] q,
\end{eqnarray}
where $q$ and $\bar{q}$ are the $u$ and $d$ quark fields with constituent quark mass $m = 270$ MeV; $\lambda$ are linear combinations of the Gell-Mann matrices \cite{Volkov:2022jfr}.
	
	The coupling constants are expressed in terms of logarithmic divergent integrals:
	\begin{eqnarray}
	\label{Couplings}
		g_{\rho} =g_{a_1}= \sqrt{\frac{3}{2I_{2}}}, &\quad& g_{\pi} = \sqrt{\frac{Z_{\pi}}{4 I_{2}}},
	\end{eqnarray}
	where
	\begin{eqnarray}
		Z_{\pi} & = & \left(1 - 6\frac{m^{2}}{M^{2}_{a_{1}}}\right)^{-1}.
	\end{eqnarray}
	Here $Z_{\pi}$ is an additional renormalization constant appearing in $\pi - a_{1}$ transitions, $M_{a_{1}} = 1230 \pm 40$~MeV is the $a_1$ meson mass \cite{ParticleDataGroup:2022pth}.

	The integral arising in quark loops has the form:
	\begin{eqnarray}
		I_{2} =
		-i\frac{N_{c}}{(2\pi)^{4}}\int\frac{\Theta(\Lambda_{4}^{2} + k^2)}{(m^{2} - k^2)^{2}}
		\mathrm{d}^{4}k = \frac{N_{c}}{(4\pi)^2}\left[ \ln\left(1+ \frac{\Lambda^2_{4}}{m^2} \right) + \frac{\Lambda^2_{4}}{\Lambda^2_{4}+m^2} \right],
	\end{eqnarray}
	where $\Lambda_{4} = 1.26$~GeV is the four-dimensional cutoff parameter \cite{Volkov:1986zb}, $N_{c} = 3$ is the number of colors in QCD.

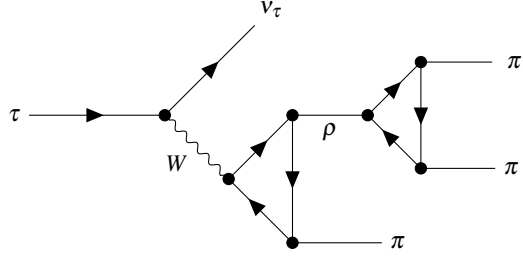
\begin{figure*}[t]
 \centering
  \begin{subfigure}{0.5\textwidth}
   \centering
    \begin{tikzpicture}
     \begin{feynman}
      \vertex (a) {\(\tau\)};
      \vertex [dot, right=2cm of a] (b){};
      \vertex [above right=2cm of b] (c) {\(\nu_{\tau}\)};
      \vertex [dot, below right=1.2cm of b] (d) {};
      \vertex [dot, above right=1.2cm of d] (e) {};
      \vertex [dot, below right=1.2cm of d] (h) {};
      \vertex [dot, right=1.0cm of e] (f) {};
      \vertex [dot, above right=1.0cm of f] (n) {};  
      \vertex [dot, below right=1.0cm of f] (m) {};   
      \vertex [right=1.2cm of n] (l) {\(\ \pi \)}; 
      \vertex [right=1.2cm of m] (s) {\(\pi\)};  
      \vertex [right=1.4cm of h] (k) {\(\pi\)}; 
      \diagram* {
         (a) -- [fermion] (b),
         (b) -- [fermion] (c),
         (b) -- [boson, edge label'=\(W\)] (d),
         (d) -- [fermion] (e),  
         (e) -- [fermion] (h),
         (d) -- [anti fermion] (h),
         (e) -- [edge label'=\({ \rho } \)] (f),
         (f) -- [fermion] (n),
         (n) -- [fermion] (m),
         (f) -- [anti fermion] (m), 
         (h) -- [] (k),
         (n) -- [] (l),
	 (m) -- [] (s),
      };
     \end{feynman}
    \end{tikzpicture}
  \end{subfigure}%
 \caption{Quark diagram for the $\tau \to 3\pi \nu_\tau$ decay contact channel.}
 \label{diagram1}
\end{figure*}

\begin{figure*}[t]
 \centering
 \centering
 \begin{subfigure}{0.5\textwidth}
  \centering
   \begin{tikzpicture}
    \begin{feynman}
      \vertex (a) {\(\tau\)};
      \vertex [dot, right=2cm of a] (b){};
      \vertex [above right=2cm of b] (c) {\(\nu_{\tau}\)};
      \vertex [dot, below right=1.2cm of b] (d) {};
      \vertex [dot, right=0.8cm of d] (l) {};
      \vertex [dot, right=1.5cm of l] (g) {};
      \vertex [dot, above right=1.2cm of g] (e) {};
      \vertex [dot, below right=1.2cm of g] (h) {};      
      \vertex [dot, right=1.0cm of e] (f) {};
      \vertex [dot, above right=1.0cm of f] (n) {};
      \vertex [dot, below right=1.0cm of f] (m) {};
      \vertex [right=1.0cm of n] (s) {\( \pi \)};
      \vertex [right=1.0cm of m] (r) {\( \pi \)};
      \vertex [right=1.2cm of h] (k) {\( \pi \)}; 
      \diagram* {
         (a) -- [fermion] (b),
         (b) -- [fermion] (c),
         (b) -- [boson, edge label'=\(W\)] (d),
         (d) -- [fermion, inner sep=1pt, half left] (l),
         (l) -- [fermion, inner sep=1pt, half left] (d),
         (l) -- [edge label'=\({ a_1, \pi } \)] (g),
         (g) -- [fermion] (h),  
         (h) -- [fermion] (e),
         (e) -- [fermion] (g),      
         (e) -- [edge label'=\( \rho \)] (f),
         (f) -- [fermion] (n),
         (n) -- [fermion] (m),
         (m) -- [fermion] (f),
         (h) -- [] (k),
         (n) -- [] (s),
         (m) -- [] (r),
      };
     \end{feynman}
    \end{tikzpicture}
  \end{subfigure}%
 \caption{Diagram with intermediate mesons describing the decay $\tau \to 3 \pi \nu_\tau$.}
 \label{diagram2}
\end{figure*}
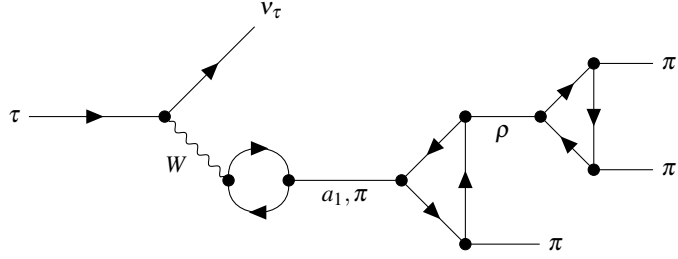

Diagrams describing the decay $\tau \to \pi^+ \pi^- \pi^- \nu_\tau$ are shown in Figures~\ref{diagram1} and \ref{diagram2}. The loops in these diagrams are formed by quark lines.

The decay amplitude of $\tau \to \pi^+ \pi^- \pi^- \nu_\tau$ takes the form:
\begin{eqnarray}
\label{amplitude1}
\mathcal{M}(\tau \to \pi^+ \pi^- \pi^- \nu_\tau) & = & -i F_{\pi} G_{F} V_{ud} L_{\mu} \left\{ \mathcal{M}_{a_1} + \mathcal{M}_{\pi} \right\}_{\mu},
\end{eqnarray}
where $G_{F}$ is the Fermi constant; $V_{ud}$ is the element of the Cabibbo-Kobayashi-Maskawa matrix; $L_{\mu}$ is lepton current; The hadronic part of the amplitude includes the axial-vector and pseudoscalar contributions:
\begin{eqnarray}
\mathcal{M}_{a_1}^{\mu} & = & Z_\pi g^2_\rho
 \left[ g_{\mu\nu} + \left((q^{2} - 6m^2)g^{\mu\nu} - q^{\mu}q^{\nu} \right) BW_{a_1} \right] BW_{\rho} (p_{\pi^+} - p^{(1)}_{\pi^-})_\nu + \left(p^{(1)}_{\pi^-} \leftrightarrow p^{(2)}_{\pi^-} \right), 
\end{eqnarray}
\begin{eqnarray}
\mathcal{M}_{\pi}^{\mu} & = & g^2_\rho BW_{\pi} q_\mu {\left(q+ p^{(2)}_{\pi^-} \right)}_\nu {\left(p_{\pi^+} - p^{(1)}_{\pi^-} \right)}_\nu BW_{\rho} + \left(p^{(1)}_{\pi^-} \leftrightarrow p^{(2)}_{\pi^-} \right), 
\end{eqnarray}
where $p_{\pi^+}$, $p^{(1)}_{\pi^-}$ and $p^{(2)}_{\pi^-}$ are the momentum of charged pions in the final state; $q$ is the momentum of intermediate axial-vector or pseudoscalar resonances. Quark loops are calculated using the methods developed in the NJL model and successfully tested on other physical processes \cite{Volkov:2005kw, Volkov:2022jfr}. The loop integrals are expanded in terms of the external fields momentum, and only the logarithmic divergent parts are preserved. Accounting for such terms makes it possible to preserve the chiral symmetry in the model \cite{Volkov:1986zb}.

Intermediate resonances are described by the Breit-Wigner propagators:
\begin{eqnarray}
BW_{M} = \frac{1}{M_{M}^{2} - q^{2} - i\sqrt{q^2}\Gamma_{M}},
\end{eqnarray}
where $M$ designates a meson. Masses and widths of mesons are taken from PDG \cite{ParticleDataGroup:2022pth}.

In the considered processes, the box diagrams have not been taken into account because their contributions
are small. A similar situation took place in the case of the decay $\tau \to K \pi\pi\nu_\tau$ in our previous paper \cite{K:2023kgj}.

Using the amplitude (\ref{amplitude1}) one can obtain numerical estimates for the branching fractions of $\tau \to 3\pi\nu_\tau$ decays. The total partial decay width of $\tau \to \pi^+ \pi^- \pi^- \nu_\tau$ calculated in the NJL model for the case $M_{a_{1}} = 1230$ MeV and $\Gamma_ {a_{1}} = 270$ MeV
\begin{eqnarray}
Br(\tau \to \pi^+ \pi^- \pi^- \nu_\tau)_{NJL} = 9.05 \, \%.
\end{eqnarray}
	
Similar calculations can be made for the decay with the production of neutral pions $\tau \to 2\pi^0\pi^-\nu_\tau$. For this process, the partial width is calculated using the amplitude (\ref{amplitude1}) with the  replacement of mass $M_{\pi^-} \to M_{\pi^0}$ in the final state. As a result, within the NJL model, we obtain the following estimate for the branching fraction:
\begin{eqnarray}
Br(\tau \to \pi^- 2\pi^0\nu_\tau)_{NJL} = 9.15 \, \%.
\end{eqnarray}

The experimental data for the branching fractions given in PDG and obtained by the Belle and Babar collaborations are:
\begin{eqnarray}
Br(\tau \to \pi^+ \pi^- \pi^- \nu_\tau) = 9.02 \pm 0.05 \, \% \, \text{\cite{ParticleDataGroup:2022pth}}, \\
Br(\tau \to \pi^+ \pi^- \pi^- \nu_\tau) = 8.83 \pm 0.13 \, \% \, \text{\cite{BaBar:2007chl}}, \\
Br(\tau \to \pi^+ \pi^- \pi^- \nu_\tau) = 8.42 \pm 0.25 \, \% \, \text{\cite{Belle:2010fal}},\\ 
Br(\tau \to \pi^- \pi^0 \pi^0 \nu_\tau) = 9.26 \pm 0.10 \, \% \, \text{\cite{ParticleDataGroup:2022pth}}.
\end{eqnarray}

As noted above, the $\tau \to 3\pi \nu_\tau$ decay was previously described in papers \cite{GomezDumm:2003ku, Dumm:2009va} in the framework of the chiral perturbative theory with resonances. In these papers, agreement with the experimental decay widths was achieved by choosing $a_1$ meson mass $M_{a_1}=1120$ MeV. This value of the $a_1$ meson mass turns out to be much lower than the experimental values \cite{ParticleDataGroup:2022pth, COMPASS:2018uzl, JPAC:2018zwp}.

Our calculations in the NJL model show the dominant contribution of the axial-vector channel in determining branching fractions of $\tau \to 3\pi \nu_\tau$.  The axial-vector channel separate contribution together with the contact channel is $Br(\tau \to a_1 \to \pi^+ \pi^- \pi^- \nu_\tau) = 8.96 \%$. The pseudoscalar channel for this process gives $Br(\tau \to \pi \to \pi^+ \pi^- \pi^- \nu_\tau) = 0.34 \%$. Negative interference between these channels is established, which reduces the total contribution of the axial vector and pseudoscalar channels to $Br(\tau \to \pi^+ \pi^- \pi^- \nu_\tau) = 9.05 \%$. It should be noted that the results in the NJL model were obtained without using additional arbitrary parameters.

In the present paper, we show that the $\tau \to 3\pi \nu_\tau$ decay width is quite satisfactorily described within our version of the NJL model when an axial-vector $a_1$ meson with mass $M_{a_1}=1230$ MeV and width $\Gamma_{a_1}=270$ MeV is used as an intermediate state. It is shown that the value of branching fractions of $\tau \to 3\pi \nu_\tau$ strongly depends on the choice of the meson width $a_1$.
	
\section*{Acknowledgments}
The authors are grateful to A.~B.~Arbuzov and A.~A.~Osipov for useful discussions.



\begin{thebibliography}{99}

\bibitem{ParticleDataGroup:2022pth}
R.~L.~Workman \textit{et al.} [Particle Data Group],
PTEP \textbf{2022} (2022), 083C01

\bibitem{COMPASS:2018uzl}
M.~Aghasyan \textit{et al.} [COMPASS],
Phys. Rev. D \textbf{98} (2018) no.9, 092003

\bibitem{JPAC:2018zwp}
M.~Mikhasenko \textit{et al.} [JPAC],
Phys. Rev. D \textbf{98} (2018) no.9, 096021

\bibitem{dArgent:2017gzv}
P.~d'Argent, N.~Skidmore, J.~Benton, J.~Dalseno, E.~Gersabeck, S.~Harnew, P.~Naik, C.~Prouve and J.~Rademacker,
JHEP \textbf{05} (2017), 143

\bibitem{LHCb:2017swu}
R.~Aaij \textit{et al.} [LHCb],
Eur. Phys. J. C \textbf{78} (2018) no.6, 443

\bibitem{Girlanda:1999fu}
L.~Girlanda and J.~Stern,
Nucl. Phys. B \textbf{575} (2000), 285-312

\bibitem{GomezDumm:2003ku}
D.~Gomez Dumm, A.~Pich and J.~Portoles,
Phys. Rev. D \textbf{69} (2004), 073002

\bibitem{Wagner:2008gz}
M.~Wagner and S.~Leupold,
Phys. Rev. D \textbf{78} (2008), 053001

\bibitem{Dumm:2009va}
D.~G.~Dumm, P.~Roig, A.~Pich and J.~Portoles,
Phys. Lett. B \textbf{685} (2010), 158-164

\bibitem{Nugent:2013hxa}
I.~M.~Nugent, T.~Przedzinski, P.~Roig, O.~Shekhovtsova and Z.~Was,
Phys. Rev. D \textbf{88} (2013), 093012

\bibitem{Sadasivan:2021emk}
D.~Sadasivan, A.~Alexandru, H.~Akdag, F.~Amorim, R.~Brett, C.~Culver, M.~D\"oring, F.~X.~Lee and M.~Mai,
Phys. Rev. D \textbf{105} (2022) no.5, 054020

\bibitem{Nambu:1961tp}
Y.~Nambu and G.~Jona-Lasinio,
Phys. Rev. \textbf{122} (1961), 345-358
doi:10.1103/PhysRev.122.345

\bibitem{Ebert:1982pk}
D.~Ebert and M.~K.~Volkov,
Z. Phys. C \textbf{16} (1983), 205
doi:10.1007/BF01571607

\bibitem{Ebert:1985kz}
D.~Ebert and H.~Reinhardt,
Nucl. Phys. B \textbf{271} (1986), 188-226
doi:10.1016/S0550-3213(86)80009-8

\bibitem{Volkov:1986zb}
M.~K.~Volkov,
Sov. J. Part. Nucl. \textbf{17} (1986), 186

\bibitem{Vogl:1991qt}
U.~Vogl and W.~Weise,
Prog. Part. Nucl. Phys. \textbf{27} (1991), 195-272
doi:10.1016/0146-6410(91)90005-9

\bibitem{Klevansky:1992qe}
S.~P.~Klevansky,
Rev. Mod. Phys. \textbf{64} (1992), 649-708
doi:10.1103/RevModPhys.64.649

\bibitem{Volkov:1993jw}
M.~K.~Volkov,
Phys. Part. Nucl. \textbf{24} (1993), 35-58

\bibitem{Ebert:1994mf}
D.~Ebert, H.~Reinhardt and M.~K.~Volkov,
Prog. Part. Nucl. Phys. \textbf{33} (1994), 1-120
doi:10.1016/0146-6410(94)90043-4

\bibitem{Volkov:2005kw}
M.~K.~Volkov and A.~E.~Radzhabov,
Phys. Usp. \textbf{49} (2006), 551-561
doi:10.1070/PU2006v049n06ABEH005905
[arXiv:hep-ph/0508263 [hep-ph]].

\bibitem{Volkov:2017arr}
M.~K.~Volkov and A.~B.~Arbuzov,
Phys. Usp. \textbf{60} (2017) no.7, 643-666
doi:10.3367/UFNe.2016.11.037964

\bibitem{Volkov:2022jfr}
M.~K.~Volkov, A.~A.~Pivovarov and K.~Nurlan,
Symmetry \textbf{14} (2022) no.2, 308
doi:10.3390/sym14020308
[arXiv:2201.03951 [hep-ph]].

\bibitem{Ivanov:1989qw}
Y.~P.~Ivanov, A.~A.~Osipov and M.~K.~Volkov,
Z. Phys. C \textbf{49} (1991), 563-568

\bibitem{K:2023kgj}
Volkov.~M.~K., Pivovarov.~A.~A. and Nurlan.~K,
[arXiv:2303.02730 [hep-ph]].

\bibitem{BaBar:2007chl}
B.~Aubert \textit{et al.} [BaBar],
Phys. Rev. Lett. \textbf{100} (2008), 011801

\bibitem{Belle:2010fal}
M.~J.~Lee \textit{et al.} [Belle],
Phys. Rev. D \textbf{81} (2010), 113007

\end{thebibliography}
\end{document}